\documentclass[fleqn,preprint,showpacs,preprintnumbers,amsmath,amssymb]{revtex4}
\usepackage{graphicx}
\usepackage{dcolumn}
\usepackage{bm}

\begin{document}

\title{Spin-induced localized density excitations in quantum plasmas}
\author{M. Akbari-Moghanjoughi}
\affiliation{Azarbaijan University of
Tarbiat Moallem, Faculty of Sciences,
Department of Physics, 51745-406, Tabriz, Iran}

\date{\today}
\begin{abstract}
In this paper the dominant effect of electron inertia on the dynamics of localized density excitations is studied in a quantum plasma in the presence of electron spin effects. Using the quantum magnetohydrodynamics (QMHD) model including electron tunneling and spin polarization phenomena, it is revealed that the quantum effects such as plasma paramagnetism and diamagnetism play inevitable role on soliton existence criteria in quantum plasmas. Furthermore, it is shown that the magnetosonic localized density-excitation stability depends strongly on the quantum system dimensionality. Two distinct region of soliton stability is shown to exist depending on the value of the electron effective mass, where, the soliton amplitude variation with respect to the external magnetic field strength is quite opposite in these regions. Current findings can be important in the study of dynamical nonlinear wave features in dense laboratory or inertial-confined plasmas.
\end{abstract}

\maketitle

\section{Introduction}

Quantum plasma has received a considerable attention during the past few years because of inevitable applications in emerging miniaturization techniques \cite{haug, Markowich}. A quantum plasma is characterized as a degenerate cold ionized matter. Due to inter-fermion distances much lower than the characteristic de Broglie thermal wavelength, $h/(2\pi m k_B T)^{1/2}$ in a quantum plasma and the domination of the Pauli exclusion rule, many quantum effects such as electron-tunneling, degeneracy pressure and Landau quantization may occur \cite{landau}. Recent studies based on quantum hydrodynamics (QHD) model reveal outstanding differences in nonlinear features of quantum plasmas from that of an ordinary ones \cite{gardner, haas1, haas2, akbari1, akbari2}. More recently, QHD model has been extended to include the spin-1/2 effects \cite{Marklund1, Brodin1, Marklund2, Brodin2}. It has been shown using the extended quantum magnetohydrodynamics (QMHD) model that the electron spin may lead to a negative pressure-like effect modifying the nonlinear wave dynamics of the quantum plasma \cite{sabry, Marklund3, Marklund4, akbari3}.

It is well-known that, under the application of small magnetic fields even the simple quantum system of non-interacting electrons at low temperatures develops complex properties. The magnetization of a simple quantum Fermi-gas has terms linear in the field such as the Pauli paramagnetism and Landau diamagnetism also much complicated terms which gives rise to oscillations as inverse magnetic field, i.e. de Haas van Alphen oscillations, may occur \cite{landau}. One of interesting effects of a magnetized plasma is the role of effective electron mass which plays an essential role in the properties of semiconducting materials. It has been shown that \cite{sond} the electron directional-inertia (effective mass) may also play a definitive role in plasma magnetization changing from paramagnetism to diamagnetism or viceversa. It is the main objection of the current investigation to probe the possible effect of the effects of electron inertia and the quantum system dimensionality on the nonlinear localized density excitations in a quantum plasma. The presentation of the paper is as follows. The QMHD plasma model including the spin contribution is introduced in Sec. \ref{equations}. The Localized density solution is given in Sec. \ref{calculation} and the numerical results are presented in Sec. \ref{discussion}. Finally, the concluding remarks are given on Sec. \ref{conclusion}.

\section{QMHD Model}\label{equations}

In a perfectly conductive and completely degenerate magnetized quantum plasma one may combine the continuity equation for the electron and ion in the following form,
\begin{equation}\label{cont}
\frac{{\partial \rho_c }}{{\partial t}} + \nabla  \cdot (\rho_c {\bf{u_c}}) = 0,
\end{equation}
where, the center of mass density and velocity are defined as $\rho_c=m_e n_e + m_i n_i \sim m_i n_i$ ($m_e\ll m_i$) and $\bm{u_c}=e(m_e n_e \bm{u_e}+m_i n_i \bm{u_i})/\rho_c$, respectively. On the other hand, the momentum equation can be written as
\begin{equation}\label{mom}
\frac{{\partial {\bf{u_c}}}}{{\partial t}} + \left( {{\bf{u_c}} \cdot \nabla } \right){\bf{u_c}} = {\rho_c ^{ - 1}}\left( {{\bf{j}} \times {\bf{B}} - \nabla P_c + {{\bf{F}}_Q}} \right),
\end{equation}
in which, $P_c$ is the total center of mass pressure, $\bm{j}$ is the magnetization current and $\bm{F}_Q={\bm{F}}_B+{\bm{F}}_S$ is a collective contribution due to quantum tunneling force ${\bm{F}}_B$, and spin-polarization, ${\bm{F}}_S$ \cite{Marklund4}. Particularly, the quantum force reads as
\begin{equation}\label{QF}
{{\bf{F}}_Q} = \frac{{{\rho_c\hbar ^2}}}{{2{m_e}{m_i}}}\nabla \frac{{\Delta \sqrt {{\rho _c}} }}{{\sqrt {{\rho _c}} }} + {{\Gamma}} \nabla {{B}},
\end{equation}
where $\bm{\Gamma}$ is the magnetization vector. For a uniformly magnetized electron-ion plasma, say, $\bm{B}=B_0\bm{k}$, combination of the above equations make closed QMHD set.
It is well known that pauli spin-magnetization per unit volume is ${{\Gamma}_P} = \left( {3\mu _B^2{n_e}/2{k_B}{T_{Fe}}} \right){{B}}$ ($T_{Fe}$ being the electron Fermi-temperature) which is related to magnetization current through the relation, ${\bf{j}} = \mu _0^{ - 1}\nabla  \times ({\bf{B}} - {\mu _0}{\bf{\Gamma}})$, where, $\mu_B=e\hbar/2m_e c$ is the Bohr magneton and $\hbar$ is the normalized Plank constant. In the case of fully degenerate quantum plasma, however, a Landau demagnetization susceptibility of $\Gamma_L=-1/3\Gamma_P$ exists due to the electron orbital contribution which results in total sum of $\Gamma = \Gamma_P + \Gamma_L = n_e\mu _B^2B/{E_{Fe}}$. The generalized susceptibility of weakly magnetized Fermi-gas ($\mu_B B_0 < k_B T_{Fe}$) in the zero-temperature limit, which also includes the effective electron mass, is given in Ref. \cite{sond}
\begin{equation}\label{QP}
\chi  = \mu _B^2D({E_{Fe}})\left[ {1 - \frac{1}{3}{{\left( {\frac{{{m_e}}}{{m_e^*}}} \right)}^2}} \right],
\end{equation}
where, $m_e^{*}$ is the effective electron mass and $D({E_{Fe}}) = d n /2 E_{Fe}$ is the density of electronic states at the Fermi level with, $d$, being the system dimensionality. It is noticed that the susceptibility may be positive or negative depending on the electron effective mass ratio in Eq. (\ref{QP}), hence, the plasma may behave paramagnetic or diamagnetic. For instance, a large fractional mass ($m_e/m_e^*\simeq 10^3$) has been reported for bismuth \cite{zang, yadong}, which gives rise to interesting quantum properties of bismuth nanotubes. On the other hand, the degeneracy pressure of a zero-temperature Fermi-gas is \cite{akbari4}
\begin{equation}\label{p}
{P_d} = \frac{{2E_{Fe }}{{n_0}}}{{(d + 2)}}{{n_e }^{\left(\frac{{d + 2}}{d}\right)}},
\end{equation} Now, assuming the propagation of magnetoacoustic nonlinear wave to be in $x$ direction perpendicular to a uniform magnetic field $B_0$ along the $z$-axis and using the quasineutrality condition $n_i\simeq n_e \simeq n$, the normalized QMHD fluid equation set, reads as
\begin{equation}\label{diff}
\begin{array}{l}
\frac{{\partial u_c}}{{\partial t}} + \frac{{\partial nu_c}}{{\partial x}} = 0, \\
{H^2}\frac{\partial }{{\partial x}}\left( {\frac{1}{{\sqrt n }}\frac{{{\partial ^2}\sqrt n }}{{\partial {x^2}}}} \right) = \frac{{\partial {u_c}}}{{\partial t}} + \frac{1}{2}\frac{{\partial u_c^2}}{{\partial x}} + \frac{{\partial {n^{2/d}}}}{{\partial x}} - \frac{{d{\epsilon^2}}}{2}\left( {1 - \frac{{{\mu ^2}}}{3}} \right)\frac{\partial }{{\partial x}}\ln n,\\
\end{array}
\end{equation}
where, we have introduced new fractional plasma entities such as the quantum diffraction parameter, $H = \sqrt {{m_i}/{2m_e}}\hbar {\omega _{pi}}/E_{Fe}$ \cite{haas2} (with $E_{Fe}$ being the electron Fermi-energy), normalized Zeeman energy, $\epsilon=\mu_B B_0/E_{Fe}$ and fractional effective mass $\mu=m_e/m_e^{*}$. Note that, in obtaining the dimensionless equations we have made use of the following scalings
\begin{equation}
x \to \frac{{{c_{s}}}}{{{\omega _{pi}}}}\bar x,\hspace{3mm}t \to \frac{{\bar t}}{{{\omega _{pi}}}},\hspace{3mm}\rho_{c} \to \bar \rho_{c}{\rho_0},\hspace{3mm}u_c \to \bar u_c{c_{s}},\hspace{3mm}\phi  \to \bar \phi \frac{{{E_{Fe }}}}{e},
\end{equation}
where, ${\omega _{pi }} = \sqrt {4\pi{e^2}n_{0}/{m_i}}$ and ${c_{s}} = \sqrt {{2E_{Fe }}/{m_i }}=v_{Fe}\sqrt{m_e/m_i}$ are characteristic plasma-frequency and the scaled electron Fermi-speed, respectively. It should be noted that, our scaling depends on dimensionality since,
\begin{equation}
{E_{Fe}} = \left\{ {\begin{array}{*{20}{c}}
{\begin{array}{*{20}{c}}
{\frac{{{\hbar ^2}}}{{2{m_e}}}\left( {\frac{\pi }{2}{n_e}} \right)} & {(d = 1)}  \\
\end{array}}  \\
{\begin{array}{*{20}{c}}
{\frac{{{\hbar ^2}}}{{2{m_e}}}{{\left( {2\pi {n_e}} \right)}^2}} & {(d = 2)}  \\
\end{array}}  \\
{\begin{array}{*{20}{c}}
{\frac{{{\hbar ^2}}}{{2{m_e}}}{{\left( {3{\pi ^2}{n_e}} \right)}^{\frac{2}{3}}}} & {(d = 3)}  \\
\end{array}}  \\
\end{array}} \right\}
\end{equation}
hence, one realizes that the quantum dispersion-relation (given below) which is obtained via Fourier analysis of Eqs. (\ref{diff}) is also a function of the system dimensionality ($d$)
\begin{equation}
\frac{\omega }{k} = \sqrt {\frac{{12 + 3d{H^2}{k^2} + {d^2}{\varepsilon^2}({\mu ^2} - 3)}}{{6d}}}.
\end{equation}
Therefore, the normalized quantum linear wave-speed, from which the Mach-number is defined (e.g. see Ref. \cite{dubinov}), is obtained as
\begin{equation}
v_{ql}=\frac{\omega }{k}\mid_{k\rightarrow0} = \sqrt {\frac{{12 + {d^2}{\varepsilon^2}({\mu ^2} - 3)}}{{6d}}}.
\end{equation}

\section{Localized Density Excitations}\label{calculation}

Coordinate transformation to $\xi=x-M t$, where, $M=V/c_s$ is the normalized matching speed of the nonlinear wave, yields the desired stationary wave solution to Eq. (\ref{diff}). By using a change of variable variable, $n=N^2$, in Eqs. (\ref{diff}) we reduced (after integration with boundary conditions $\mathop {\lim }\limits_{\xi  \to  \pm \infty } n = 1$ and $\mathop {\lim }\limits_{\xi  \to  \pm \infty } u_c = 0$) to a single differential equation of the form
\begin{equation}\label{diff2}
\frac{{{H^2}}}{N}\frac{{{\partial ^2}N}}{{\partial {\xi ^2}}} = \frac{{{M^2}}}{2}{\left( {1 - {N^{ - 2}}} \right)^2} - {M^2}\left( {1 - {N^{ - 2}}} \right) + ({N^{4/d}} - 1) - {{d{\epsilon^2}}}\left( {1 - \frac{{{\mu ^2}}}{3}} \right)\ln N,
\end{equation}
An energy-like relation ${({d_\xi }n)^2}/2 + U(n) = 0$ can be obtained multiplying both sides of Eq. (\ref{diff2}) with $dN/d\xi$ and integrating with aforementioned boundary conditions in which the desired pseudopotential is given as
\begin{equation}\label{pseudo}
\begin{array}{l}
U(n) = \frac{1}{{3(d+2){H^2}}}\left\{ {n\left[ {(d+2)\left( {6n + d{\epsilon^2}({\mu ^2-3})(n - 1)} \right)} \right.} \right.\left. { - 12 - 6d{n^{(d + 2)/d}}} \right] \\ \left. { + 3{M^2}(d+2){{(n - 1)}^2} - {n^2}{\epsilon^2}d(d+2)({\mu ^2} - 3)\ln n} \right\}. \\
\end{array}
\end{equation}
It is confirmed that the pseudopotential and its first derivative vanishes at $n=1$. In order to evaluate the possibility of solitary excitations we further evaluate the second derivative of the pseudopotential at the unstable point $n=1$ which gives rise to
\begin{equation}\label{dd}
{\left. {\frac{{{d^2}U(n)}}{{d{n^2}}}} \right|_{n = 1}} = \frac{{12 + 6d{M^2} - {d^2}{\epsilon^2}({\mu ^2} - 3)}}{{3d{H^2}}},
\end{equation}
hence, in general we have
\begin{equation}\label{conditions}
{\left. {U(n)} \right|_{n = 1}} = {\left. {\frac{{dU(n)}}{{dn}}} \right|_{n = 1}} = 0,\hspace{3mm}{\left. {\frac{{{d^2}U(n)}}{{d{n^2}}}} \right|_{n = 1}} < 0.
\end{equation}
Finally we require that for at least one either maximum or minimum nonzero $n$-value, we have $U(n_{m})=0$, so that for every value of $n$ in the range ${n _m} > n  > 1$ (compressive soliton) or ${n _m} < n  < 1$ (rarefactive soliton), $U(n)$ is negative (it is understood that there is no root in the range $[1,n_m]$). In such a condition we can obtain a potential minimum which describes the possibility of a solitary wave propagation. The stationary soliton solutions corresponding to this pseudopotential which satisfies the mentioned boundary-conditions, then, read as
\begin{equation}\label{soliton}
\xi  - {\xi _0} =  \pm \int_1^{n_m} {\frac{{dn}}{{\sqrt { - 2U(n)} }}}.
\end{equation}
Therefore, for the existence of a solitary excitation the corresponding matching speed value should be below a critical value. i.e.
\begin{equation}\label{consol}
M < M_{cr} = {\sqrt {\frac{{12 + {d^2}{\epsilon^2}({\mu ^2} - 3)}}{{6d}}} }.
\end{equation}
Now, we evaluate the existence of $n_m$ values which is essential to our analysis. A close inspection of the pseudopotential given in Eq. (\ref{pseudo}) reveals that
\begin{equation}\label{nm}
\mathop {\lim }\limits_{n \to 0} U(n) =  \frac{{{M^2}}}{{{H^2}}} > 0,\hspace{3mm}\mathop {\lim }\limits_{n \to  + \infty } U(n) =  - \infty,
\end{equation}
which is the indication of the fact that only rarefactive density excitations may exist in this plasma regardless of all plasma fractional parameters. These findings are quite similar to the one reported for unmagnetized plasma case in Ref. \cite{akbari4}

\section{Numerical Interpretations}\label{discussion}

In Fig. 1 the volume (in $M$-$\epsilon$-$\mu$ space) in which a localized density structure can occur is shown. The left column (Figs. 1(a)-1(c)) depicts the variation of the corresponding volume with the variation in quantum dimensionality when the fractional electron mass, $\mu$ is below the critical value of $\sqrt{3}$. It is clearly evident that the increase in the dimensionality reduces the volume of soliton stability region. For all dimensionality values it is observed that below the critical fractional electron mass the increase in the strength of the magnetic field, $B_0$ (or correspondingly increase in the normalized Zeeman-energy, $\epsilon$) leads to a decrease in the critical soliton matching-speed value. However, this decrease gets sharper as the dimensionality of the quantum system increases. It is also noted that, the increase in the fractional electron mass for a fixed \emph{non-zero} magnetic field strength causes the volume of soliton stability to expand to higher M-values. It should be noted that the shown range of variation for $\epsilon$ in this figure is somehow exaggerated in order to clarify the effect, since, this value should be much smaller than unity. In the right column, i.e. Fig. 1(d)-1(f), the soliton stability volume is shown for $\mu>\sqrt{3}$-values, which clearly indicate a different behavior. Although in this case the for small $\epsilon$ and $\mu$ values the soliton occurrence volume with the system dimensionality is similar to the cases of $\mu<\sqrt{3}$, however, for the corresponding large values of $\mu$ this effect is completely reversed. More importantly, in the case of $\mu>\sqrt{3}$, for all system dimensionality, the increase in $\epsilon$, (or equivalently the strength of the magnetic field) causes the increase in the critical M-values, opposite to the case shown in Figs. 1(a)-1(c). However, for $\mu>\sqrt{3}$ the increase in the value of $\mu$, similar to the case of $\mu<\sqrt{3}$ expand the solitons occurrence volume to higher $M$-values.

Figure 2(a)-2(f) show the variations of pseudopotential with changes in each of the values of soliton matching-speed, $M$, normalized Zeeman energy, $\epsilon$, fractional electron effective inertia, $\mu$, and the quantum diffraction parameter, $H$, when other parameters are fixed. The obtained solitary excitation profiles for values of parameters used in Fig. 2 is shown in Figs. 3(a)-3(f). Comparing Figs. 3(a) and 3(b) reveals that the soliton amplitude decreases when $\mu<\sqrt{3}$ while it increases as $\mu>\sqrt{3}$. This effect marks a dominant role of electron inertia in soliton dynamics in magnetized quantum plasmas. On the other hand, Fig. 3(c) indicates that variation of the quantum diffraction effect has absolutely no effect on the soliton amplitude and only change the solitary profile width. This feature is quite consistent with the findings of Ref. \cite{akbari4} for an unmagnetized quantum plasma. Also one should notice that variation of soliton profile with change of system dimensionality in Fig. 3(d) which is in complete agreement with finding of the aforementioned reference. Figure 3(e) reveals that as the fractional effective electron mass ($m_e/m_e^{*}$) increases, the soliton amplitude is also increased. Figure 3(f) reveals that the faster solitons are smaller in amplitude and wider in width in this plasma, which is also in agreement with findings of Ref. \cite{akbari4}.

\section{Conclusions}\label{conclusion}

Using the conventional Sagdeev potential method in the framework of spin-included magnetohydrodynamics model, it was shown that the critical electron effective mass exists in a magnetized Fermi-plasma below and above which the effect of magnetic field strength on propagation of magnetoacoustic waves is quite opposite. It has also been revealed that other plasma parameters such as soliton matching-speed, normalized Zeeman-energy, quantum diffraction parameter and fractional electron inertia has significant effects on the localized density excitations in a magnetized quantum plasma. The findings presented in this article can be helpful in the dynamical properties of dense laboratory plasmas such as semiconductors, nano-fabricated compounds and inertial-confined plasmas.

\newpage

\textbf{FIGURE CAPTIONS}

\bigskip

Figure-1

\bigskip

The volume (in $M$-$\epsilon$-$\mu$ space) in which a localized density structure can occur is shown. The left/right column depicts the variation of the corresponding volume with the variation in quantum dimensionality when the fractional electron mass, $\mu$ is below/above the critical value of $\sqrt{3}$.

\bigskip

Figure-2

\bigskip

The variations of pseudopotential depth and width for rarefactive solitary spin-polarized nonlinear magnetosonic waves with respect to change in each of three independent plasma fractional parameter, normalized soliton-speed, $M$, normalized Zeeman-energy, $\epsilon$, the quantum diffraction parameter, $H$, and the fractional effective electron-mass, $\mu$, while the other three parameters are fixed. The dash-size of the potential curves increase with regard to increase in the varied parameter.

\bigskip

Figure-3

\bigskip

The variations of localized magnetoacoustic density-excitation profiles with respect to change in each of three independent plasma fractional parameter, normalized soliton-speed, $M$, normalized Zeeman-energy, $\epsilon$, the quantum diffraction parameter, $H$, and the fractional effective electron-mass, $\mu$, while the other three parameters are fixed. The thickness of the profile increase with regard to increase in the varied parameter.

\bigskip

\newpage

\begin{figure}[ptb]\label{Figure1}
\includegraphics[scale=.5]{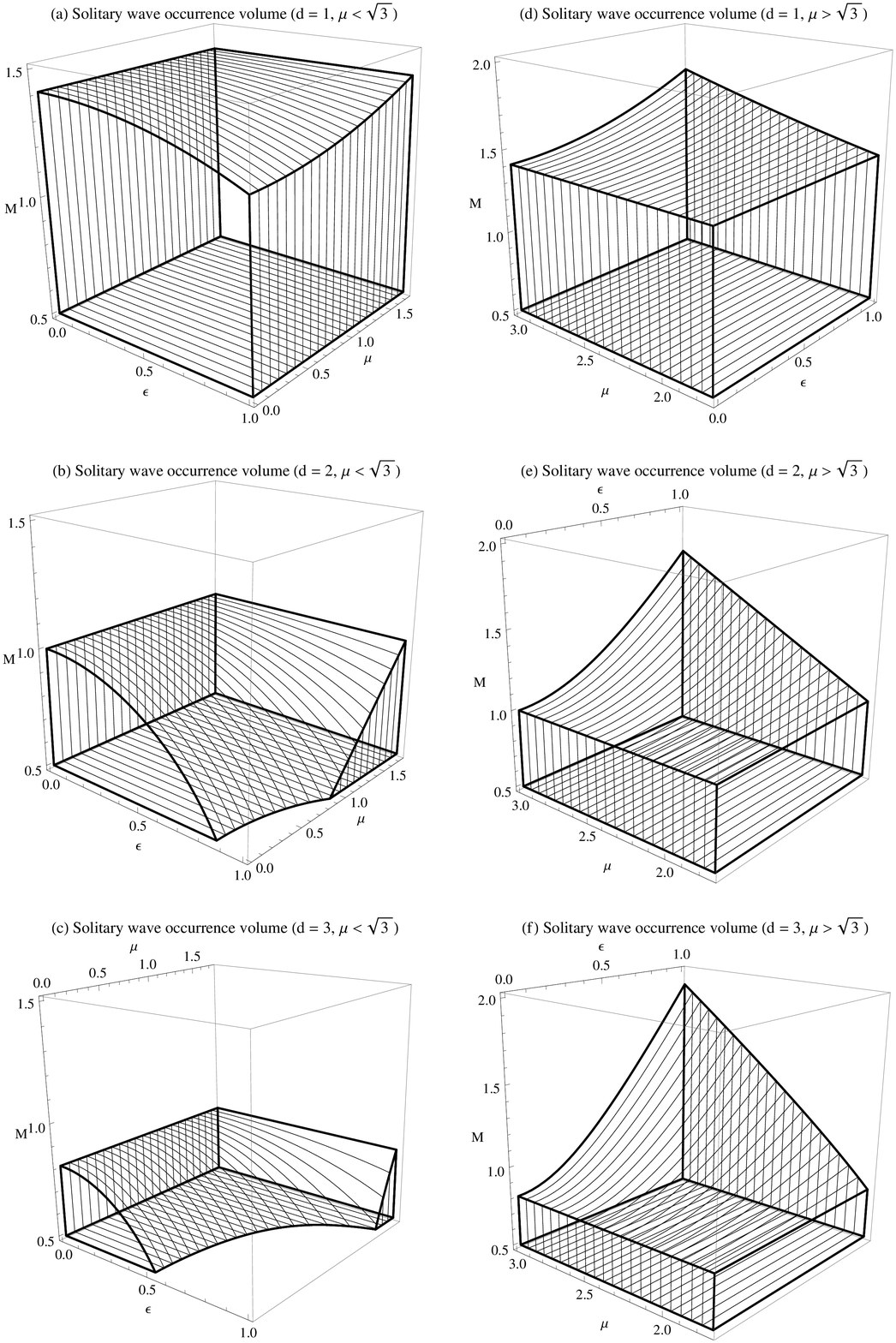}\caption{}
\end{figure}

\newpage

\begin{figure}[ptb]\label{Figure2}
\includegraphics[scale=.5]{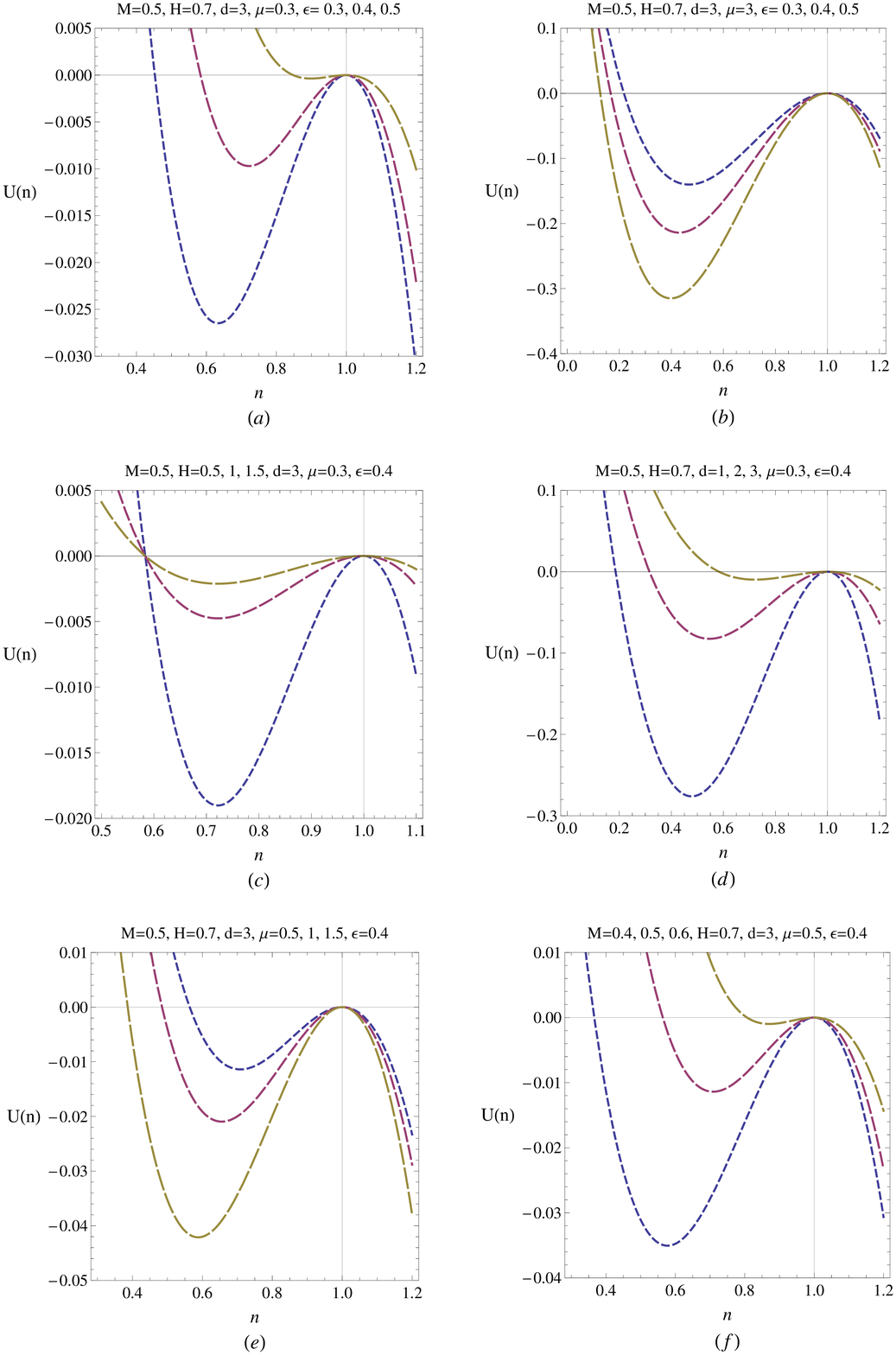}\caption{}
\end{figure}

\newpage

\begin{figure}[ptb]\label{Figure3}
\includegraphics[scale=.5]{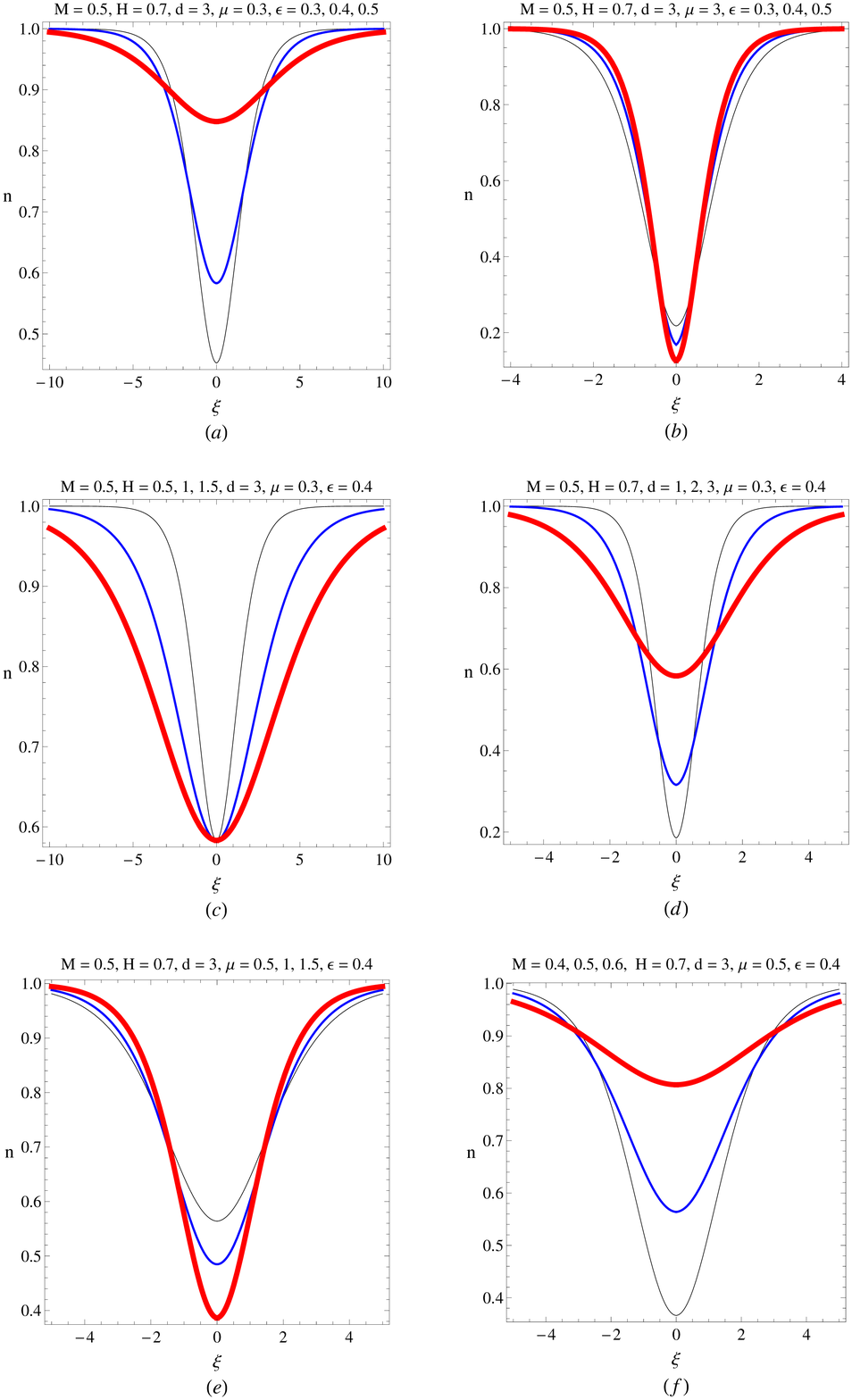}\caption{}
\end{figure}

\end{document}